\documentclass[12pt]{article}
\usepackage{graphicx,a4,epsfig,here,rotate,amsmath}
\usepackage{latexsym,color,citesort,amssymb}
\usepackage{booktabs}

\newcommand{\be}{\begin{equation}}
\newcommand{\ee}{\end{equation}}
\newcommand{\eq}{\begin{eqnarray}}
\newcommand{\en}{\end{eqnarray}}
\newcommand{\bea}{\begin{eqnarray}}
\newcommand{\eea}{\end{eqnarray}}

\newcommand{\ed}{\end{document}}
\newcommand{\bc}{\begin{center}}
\newcommand{\ec}{\end{center}}

\begin{document}

\thispagestyle{empty}
\begin{center}

\vspace{3cm}
{\Large{\bf Three-particle bound states in a finite volume:\\[0.2cm]
Unequal masses and higher partial waves}}

\vspace{0.5cm}
\today

\vspace{0.5cm}

Yu Meng$^{a}$,
Chuan Liu$^{a,b}$,
Ulf-G. Mei{\ss}ner$^{c,d}$ and
A. Rusetsky$^c$

\vspace{1em}

\begin{tabular}{c}
$^a\,${\it School of Physics and Center for High Energy Physics,}\\
{\it Peking University, Beijing 100871, P.R. China}\\[2mm]
$^b\,${\it Collaborative Innovation Center of Quantum Matter,}\\
{\it Beijing 100871, P.R. China}\\[2mm]
$^c\,${\it Helmholtz--Institut f\"ur Strahlen-- und Kernphysik and}\\
{\it Bethe Center for Theoretical Physics,}\\
{\it Universit\"at Bonn, D--53115 Bonn, Germany}\\[2mm]
$^d${\it Institute for Advanced Simulation (IAS-4),
Institut f\"ur Kernphysik  (IKP-3),}\\
{\it J\"ulich Center for Hadron Physics and JARA-HPC}\\
{\it Forschungszentrum J\"ulich, D-52425 J\"ulich, Germany}\\[2mm]

\end{tabular}

\end{center}

\vspace{1cm}

{\abstract
An explicit expression for the finite-volume energy shift of shallow three-body
bound states for non-identical particles is obtained in the
unitary limit. The inclusion of the higher partial waves is considered. To this end,
the method of Ref.~\cite{MRR} is generalized for the case of unequal masses and
arbitrary angular momenta. It is shown that in the S-wave and in the equal mass limit, the
result from Ref.~\cite{MRR} is reproduced.

\vspace*{1.cm}

}

\clearpage


\section{Introduction}

In the analysis of lattice data, the L\"uscher formalism is used
both to evaluate the finite-volume corrections to the stable particle
masses~\cite{Luescher-1}, as well as to extract the two-body scattering
lengths and scattering phase shifts
from the finite-volume energy spectra of the two-particle
systems~\cite{Luescher-2,Luescher-torus}. However, a generalization of the above
finite-volume approach from two- to three-particle case turned out to be a rather
challenging task. Only in the last few years, this issue has been addressed extensively
in the literature~\cite{Polejaeva:2012ut,Briceno:2012rv,Hansen-Sharpe,Hansen:2015zga,Hansen:2015zta,Hansen:2016fzj,Hansen-corr,Briceno:2018mlh,Mai:2017bge,Briceno:2017tce,Guo:2017crd,Guo:2017ism,Guo:2016fgl,Kreuzer:2008bi,Kreuzer:2009jp,Kreuzer:2010ti,Kreuzer:2012sr,Jansen:2015lha,Bour:2012hn,Bour:2011ef,MRR,Pang1,Pang2,Doring:2018xxx}.
Despite the significant effort, the progress has been slow so far. Namely, the finite volume
spectrum of the three-particle system in some simple models has been calculated only
very recently~\cite{Doring:2018xxx,Briceno:2018mlh} (see also earlier work~\cite{Kreuzer:2008bi,Kreuzer:2009jp,Kreuzer:2010ti,Kreuzer:2012sr}, where exclusively the three-body
bound-state sector was addressed). Such calculations are very useful since,
at this stage,
one does not yet have enough insight into the problem and lacks intuition
to predict the behavior of the three-particle finite-volume energy levels. Moreover,
these calculations might facilitate the interpretation of a particular behavior
of the energy spectrum in terms of various physical phenomena in the infinite volume.

For the reasons given above, it is very interesting to study
the few simple three-body systems, for which an analytic solution in a
finite volume is available. The three-body bound state is one of these.
In Ref.~\cite{MRR}, it has been shown that it is possible to obtain an explicit expression
for the leading order finite volume energy shift of the
S-wave shallow bound state of three identical bosons in the unitary limit, i.e.,
when the two-particle scattering length tends to infinity and the effective range
(and higher order shape parameters) are zero
(the so-called Efimov states, see Ref.~\cite{Efimov}).
 This expression  has a remarkably
simple form:
\eq
\frac{\Delta E}{E_T}=c\,(\kappa L)^{-3/2}|A|^2\exp\biggl(-\frac{2\kappa L}{\sqrt{3}}\biggr)\, .
\en
In this expression, $L$ is the side length of the spatial cubic box, $E_T$ and $\Delta E$ denote the binding energy
and the shift, respectively, $\kappa=\sqrt{m_0E_T}$ is the bound state momentum ($m_0$ denotes the mass of the
particle), and $c\simeq-96.351$ is the numerical coefficient. Further, $A$ is the so-called asymptotic
normalization coefficient for the bound state (it is equal to one, if no derivative three-particle forces
are present). The formula is valid when $\kappa L\gg 1$. Later, the same formula has been obtained in
Ref.~\cite{Hansen-corr}, using the three-particle quantization condition from Ref.~\cite{Hansen-Sharpe},
and in Ref.~\cite{Pang1} by using the finite-volume particle-dimer formalism,
formulated in Refs.~\cite{Pang1,Pang2}. Moreover, in Ref.~\cite{Pang1} the role
of the three-particle force (encoded in the asymptotic normalization coefficient) has been
clarified, and the condition of an infinitely large two-body scattering length has been relaxed.
By doing this, one can nicely observe a continuous transition from the bound state of a tightly
bound dimer and a spectator to the loosely bound three particle bound state.

It should be especially mentioned that the functional $L$-dependence of the energy shift differs
from the one predicted by the two-particle L\"uscher formula~\cite{Luescher-1} (see also
Ref.~\cite{Koenig-Lee} where the $n$-particle bound state is considered), which would be the case,
when the three-particle bound state could be represented as a loosely bound state of a tightly bound
dimer and a spectator, as well as from the perturbative shift of the three-particle ground state,
which has been derived, e.g., in Refs.~\cite{Beane-Detmold-Savage,Sharpe}. In this sense, the three-body
bound state problem represents a highly non-trivial testing ground for all theories that describe
the spectrum of the three-particle system in a finite volume.

In the present paper, we generalize the original result of Ref.~\cite{MRR} to the case
of non-identical particles and include higher partial waves. This problem is interesting,
first and foremost because, to the best of our knowledge,
all available explicit results in the three-body sector so far are limited to the S-wave
states only. Carrying out benchmark calculations in higher partial waves will enable
one to carry out more elaborate tests and to understand much better the
three-particle dynamics in a finite volume that is important for analyzing
simulation data from
lattice QCD for the three-particle systems. This is exactly the aim of this short, technical
article. Eventually, it would be interesting to study the same problem in moving frames
and consider the particles with spin. This, however, forms a subject of a separate
investigation and will be addressed in the future.

\section{Derivation of the energy shift formula}

\subsection{Notations}

The wave function of three non-identical bosons obeys the Schr\"odinger equation:
\eq
\biggl\{\sum_{i=1}^3\biggl(-\frac{1}{2m_i}\,\nabla_i^2+V_i({\bf x}_i)\biggr)
+E_T\biggr\}\psi({\bf r}_1,{\bf r}_2,{\bf r}_3)=0\, ,
\en
where $\nabla_i=\partial/\partial{\bf r}_i$.
In the following, we always assume that $(ijk)$ form an even permutation,
and $i,j,k$ can take the values $1,2,3$.
Also, we mainly follow the notations and conventions of Ref.~\cite{Nielsen}.
The relative coordinates are defined as:
\eq
{\bf x}_i=\mu_{jk}({\bf r}_j-{\bf r}_k)\, ,\quad\quad
{\bf y}_i=\mu_{i(jk)}\biggl(\frac{m_j{\bf r}_j+m_k{\bf r}_k}{m_j+m_k}-{\bf r}_i\biggr)\, ,
\en
where
\eq
\mu_{jk}=\sqrt{\frac{m_jm_k}{M(m_j+m_k)}}\, ,\quad\quad
\mu_{i(jk)}=\sqrt{\frac{m_i(m_j+m_k)}{M(m_i+m_j+m_k)}}\, .
\en
Here, $M$ denotes some normalization mass. The observables do not depend
on the choice of $M$.
If $m_1=m_2=m_3=m_0$, the choice $M=m_0/2$ corresponds to the conventions of
Ref.~\cite{MRR} that makes the comparison simpler.
For this reason, we shall choose $M=(m_1+m_2+m_3)/6$ in the
following. The bound-state momentum is defined as:
\eq
E_T=\frac{\kappa^2}{2M}\, .
\en
There are three different sets of relative coordinates. The relation between them
is given by
\eq\label{eq:ij}
{\bf x}_j=-{\bf x}_i\cos\gamma_{ij}+{\bf y}_i\sin\gamma_{ij}\, ,\quad\quad
{\bf y}_j=-{\bf x}_i\sin\gamma_{ij}-{\bf y}_i\cos\gamma_{ij}\, ,
\en
where
\eq
\gamma_{ij}=\arctan\biggl(\sqrt{\frac{m_k(m_i+m_j+m_k)}{m_im_j}}\biggr)\, ,\quad
\quad
-\frac{\pi}{2}\leq\gamma_{ij}\leq\frac{\pi}{2}\, .
\en
The hyperradius $R$ and the hyperangles $\alpha_i$ are defined as:
\eq\label{eq:alphaij}
|{\bf x}_i|=R\sin\alpha_i\, ,\quad
|{\bf y}_i|=R\cos\alpha_i\, ,\quad
R^2={\bf x}_i^2+{\bf y}_i^2\, .
\en
The relation between different hyperangles is given by:
\eq\label{eq:costheta}
\sin^2\alpha_j=\sin^2\alpha_i\cos^2\gamma_{ij}+\cos^2\alpha_i\sin^2\gamma_{ij}
-2\cos\alpha_i\sin\alpha_i\cos\gamma_{ij}\sin\gamma_{ij}\cos\theta_i\, ,
\en
where $\theta_i$ is the angle between the ${\bf x}_i$ and ${\bf y}_i$.

\smallskip

The six-dimensional integration measure is written as
\eq\label{eq:hyperspherical}
d^3{\bf x}_id^3{\bf y}_i=R^5dR\sin^2\alpha_i\cos^2\alpha_id\alpha_i
d\Omega_{x_i}d\Omega_{y_i}\, ,
\en
where $\Omega_{x_i}$, $\Omega_{y_i}$ denote the solid angles in the direction
of the vectors ${\bf x}_i$ and ${\bf y}_i$, respectively.

\smallskip

The wave function, expressed in terms of the ${\bf x}_i,{\bf y}_i$, takes the form
\eq
\psi({\bf r}_1,{\bf r}_2,{\bf r}_3)=\psi^i({\bf x}_i,{\bf y}_i)\, .
\en

\subsection{The energy shift}

A straightforward generalization of the energy shift formula of Ref.~\cite{MRR} gives:
\eq\label{eq:12}
\Delta E&=&
\sum_{i=1}^3\sum_{{\bf p},{\bf q},{\bf n},{\bf l}}\sum_{{\bf k}\neq -({\bf l}+{\bf n})}\int d^3{\bf x}_id^3{\bf y}_i
\nonumber\\[2mm]
&\times&\biggl(\psi^i\bigl({\bf x}_i-({\bf p}+{\bf q})\mu_{jk}L,
{\bf y}_i+\frac{\mu_{i(jk)}L}{m_j+m_k}({\bf p}m_k-{\bf q}m_j)\bigr)\biggr)^*
V_i({\bf x}_i+\mu_{jk}{\bf k}L)
\nonumber\\[2mm]
&\times&\psi^i\bigl({\bf x}_i-({\bf n}+{\bf l})\mu_{jk}L,
{\bf y}_i+\frac{\mu_{i(jk)}L}{m_j+m_k}({\bf n}m_k-{\bf l}m_j)\bigr)\, ,
\en
where ${\bf p},{\bf q},{\bf k},{\bf l},{\bf n}\in\mathbb{Z}^3$. Note that
the periodic boundary conditions are assumed.

In order to obtain the energy shift at leading order, we use the following procedure. First,
we shift the variables
\eq
{\bf x}_i\to {\bf x}_i-\mu_{jk}{\bf k}L\, ,\quad\quad
{\bf y}_i\to {\bf y}_i-\frac{\mu_{i(jk)}L}{m_j+m_k}({\bf p}m_k-{\bf q}m_j)\, .
\en
Next, we take into account the fact that the wave function of the bound state decreases
exponentially when the hyperradius becomes large. The suppression factor is given
by $\exp(-\kappa R)$. The equation~(\ref{eq:12}) contains two wave functions
with different arguments -- we refer to them as to the first and the second
wave functions in the following. It is immediately seen that
in the sum over  ${\bf p},{\bf q},{\bf k},{\bf l},{\bf n}$
the leading contribution is given by those term(s), where the sum
of the hyperradii for the first and the second wave functions
$R_1+R_2$ is mini\-mal
as $L\to\infty$. All other terms will give contributions that are exponentially suppressed
with respect to this contribution. Writing down explicitly
\eq
R_1+R_2&\!\!=\!\!&\mu_{jk}L\biggl\{ |{\bf p}+{\bf q}+{\bf k}|+
\biggl( ({\bf n}+{\bf l}+{\bf k})^2
+\biggl(\frac{\mu_{i(jk)}}{2\mu_{jk}(m_j+m_k)}\biggr)^2
\nonumber\\[2mm]
&\!\!\times\!\!&
\bigl((m_j+m_k)(-{\bf l}+{\bf q}+{\bf n}-{\bf p})
+(m_j-m_k)(-{\bf l}+{\bf q}-{\bf n}+{\bf p})\bigr)^2\biggr)^{1/2}\biggr\}\, ,
\en
one can straightforwardly check that the following choices
\eq\label{eq:sol1}
{\bf n}+{\bf l}+{\bf k}={\bf e}\, ,\quad
{\bf p}+{\bf q}+{\bf k}={\bf 0}\, ,\quad
 -{\bf l}+{\bf q}+{\bf n}-{\bf p}=-{\bf e}\, ,
\en
and
\eq\label{eq:sol2}
{\bf n}+{\bf l}+{\bf k}={\bf e}\, ,\quad
{\bf p}+{\bf q}+{\bf k}={\bf 0}\, ,\quad
 -{\bf l}+{\bf q}+{\bf n}-{\bf p}={\bf e}\, ,
\en
where ${\bf e}$ is the unit vector with $|{\bf e}|=1$, lead to the minimum of $R_1+R_2$,
if all relevant permutations $(ijk)=(123),(231),(312)$ are considered\footnote{Note that the situation
here is rather subtle. Namely, if we consider a fixed choice of $(ijk)$, for some mass
ratios there exist solutions, other than in Eqs.~(\ref{eq:sol1},\ref{eq:sol2}),
which lead to the lower value of $R_1+R_2$. What we claim here, is that this value
of $R_1+R_2$ is still higher than the value, obtained from  Eqs.~(\ref{eq:sol1},\ref{eq:sol2}) for
another choice of $(ijk)$. In other words, we claim that Eq.~(\ref{eq:Eshift}) always contains a
leading exponential, along with some subleading pieces. On the other hand, one has to retain
these subleading pieces as well, if one wants to reproduce the result in the equal mass limit.}.
Thus, the energy shift formula simplifies to
\eq\label{eq:Eshift}
\Delta E&\!=\!&
\sum_{\bf e}\sum_{i=1}^3\int d^3{\bf x}_id^3{\bf y}_i(\psi^i({\bf x}_i,{\bf y}_i))^*V_i({\bf x}_i)
\psi^i\biggl({\bf x}_i-\mu_{jk}{\bf e}L,{\bf y}_i-\mu_{i(jk)}\frac{m_j{\bf e}L}{m_j+m_k}\biggr)
\nonumber\\[2mm]
&\!+\!&
\sum_{\bf e}\sum_{i=1}^3\int d^3{\bf x}_id^3{\bf y}_i(\psi^i({\bf x}_i,{\bf y}_i))^*V_i({\bf x}_i)
\psi^i\biggl({\bf x}_i-\mu_{jk}{\bf e}L,{\bf y}_i+\mu_{i(jk)}\frac{m_k{\bf e}L}{m_j+m_k}\biggr)\, ,
\en
where the sum runs over the six possible orientations of the unit vector ${\bf e}$.

\subsection{The wave function for a state with an arbitrary angular momentum}
From Ref.~\cite{Nielsen} one may read off the explicit form of the wave function of the
three-particle bound state in the unitary limit:
\eq\label{eq:psilm}
\psi({\bf r}_1,{\bf r}_2,{\bf r}_3)=\sum_{i=1}^{3}
\phi^i({\bf x}_i,{\bf y}_i)\, ,
\en
where, for a given orbital momentum $l$ and projection $m$,
\eq\label{eq:philm}
\phi^i_{lm}({\bf x}_i,{\bf y}_i)&=&N_{l_xl_yl}R^{-5/2}f(R)\sum_{l_xl_y}
A_i^{(l_xl_y)}\sin^{l_x}\alpha_i\cos^{l_y}\alpha_i
P_{\nu}^{\frac{1}{2}+l_x,\frac{1}{2}+l_y}(-\cos2\alpha_i)
\nonumber\\[2mm]
&\times&
\sum_{m_x+m_y=m}c^{lm}_{l_xm_x,l_ym_y}
Y_{l_xm_x}(\Omega_{x_i})Y_{l_ym_y}(\Omega_{y_i})\, .
\en
Here, the $P_\nu^{(a,b)}(x)$ denote Jacobi functions, $Y_{lm}(\Omega)$ are spherical
harmonics, the $c^{lm}_{l_xm_x,l_ym_y}$ denote the Clebsch-Gordan coefficients
and $f(R)$ is the radial function. The wave functions, which in this paper
are used in the calculation of the energy shift, obey the Bose-symmetry in
case of identical particles, see Refs.~\cite{Nielsen,Nielsen:1999vt,Navratil:1998sv} for
more details.

The three-particle bound states in the unitary limit exist only if the resonant interaction
is in an S-wave, i.e., $l_x=0$~\cite{Nielsen}. Then, $l_y=l$.
The coefficients $A_i^{(0,l)}\doteq A_i$ obey the linear equations:
\eq
\begin{pmatrix}P & Q_{12} & Q_{31}
\cr Q_{12} & P & Q_{23} \cr Q_{31} & Q_{23} & P\end{pmatrix}
\begin{pmatrix}A_1 \cr A_2 \cr A_3\end{pmatrix}=0\, ,
\en
where
\eq
P&=&\frac{\sin((\nu+\frac{3}{2})\pi)}{\sin(\frac{3}{2}\,\pi)}\, ,
\nonumber\\[2mm]
Q_{ij}=Q_{ji}&=&\frac{\Gamma(\frac{3}{2})\Gamma(\nu+\frac{3}{2}+l)}
{\Gamma(\frac{3}{2}+l)\Gamma(\nu+\frac{3}{2})}
F(-\nu,\nu+l+2,\frac{3}{2}+l,\cos^2\gamma_{ij})(-\cos\gamma_{ij})^l\, ,
\en
in terms of Gamma and hypergeometric functions.
In order to have a non-trivial solution to this homogeneous system of linear equations,
the determinant of this system must be equal to zero. This determines the discrete
values of the parameter $\nu$. One further defines
\eq
\nu=-\frac{1}{2}(2+l)+\frac{1}{2}\,\sqrt{4+\lambda}\, ,\quad\quad
\lambda=-4-\xi^2\, .
\en
The radial wave function is given by the same expression for all $l$:
\eq
f(R)=R^{1/2}K_{i\xi}(\kappa R)\, ,
\en
where $K_\mu(z)$ denotes the modified Bessel function. Bound states occur when
$\xi$ is real, i.e., when $\lambda<-4$. In the S-wave, $l=0$, this happens for all values
of the masses $m_1,m_2,m_3$. However, if $l\neq 0$, one of the masses must be much
lighter than other two, in order that  Efimov states  can emerge~\cite{Nielsen}
(see also Ref.~\cite{Helfrich}, where the properties of  Efimov states
in higher partial waves are discussed).
Consequently, the treatment of  bound states in higher partial waves is not
possible if only the equal-mass case is considered.

The wave function of a bound state is always normalized to unity. We shall in
addition assume that
\eq
\sum_{i=1}^3 A_i^2=1\, .
\en
This is equivalent to the assumption that the asymptotic normalization coefficient
$A=1$ or, equivalently, only  non-derivative three-particle interactions are present
in the system. In the following, we shall stick to this assumption.

\section{Results and discussion}

Before considering the case of arbitrary $l$, we discuss the most interesting cases $l=0,1$ in detail.

\subsection{The case $l=0$}

\begin{figure}[t]

\begin{center}
\includegraphics*[width=9.cm]{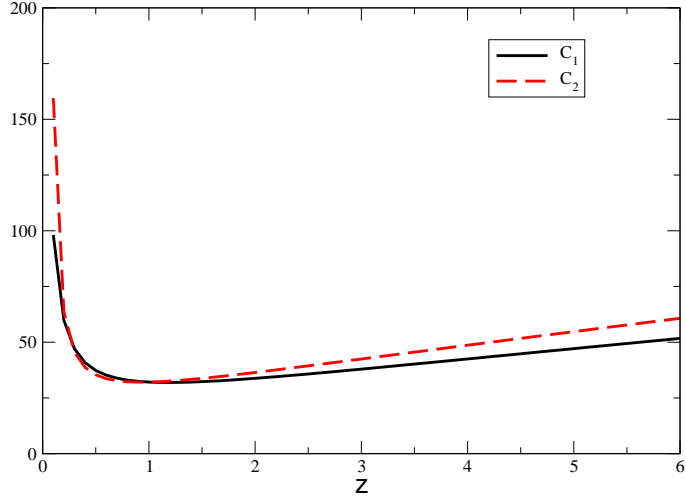}
\caption{The coefficients $C_1$ (solid line) and $C_2=C_3$ (dashed line)
as a function of the mass ratio $z=m_1/m_2=m_1/m_3$, see Eq.~(\ref{eq:Cz}).}\label{fig:Cz}
\end{center}

\end{figure}

The wave function is given by:
\eq
\psi^i_{00}({\bf x}_i,{\bf y}_i)=N_{000}\,2\sqrt{3} R^{-5/2}f(R)\sum_{i=1}^3
A_i\frac{\sinh(\xi(\frac{\pi}{2}-\alpha_i))}{\sin(2\alpha_i)}\, .
\en
Here, we have introduced an additional factor $2\sqrt{3}$ in the normalization that
allows an easier comparison with the results of Ref.~\cite{MRR}.
It is clear that, in the equal-mass case, the wave function is totally
symmetric with respect to the permutation of all particles.
Further, the wave function obeys the following condition:
\eq\label{eq:Vpsi0}
V_j({\bf x}_j)\psi^j({\bf x}_j,{\bf y}_j)=-\delta^{(3)}({\bf x}_j)F_0({\bf y}_j)\, ,
\en
where
\eq\label{eq:F0}
F_0({\bf y}_j)=N_{000}\sqrt{3}\,\frac{2\pi}{M}\,\frac{A_j}{|{\bf y}_j|}\,K_{i\xi}(\kappa |{\bf y}_j|)
\sinh\biggl(\frac{\pi\xi}{2}\biggr)\, .
\en
The normalization condition gives
\eq
N_{000}^2=\kappa^2c_0\, ,
\en
where
\eq
c_0^{-1}&=&\frac{12\pi^3\xi}{\sinh(\pi\xi)}\,
\biggl\{\biggl(\frac{1}{2\xi}\,\sinh(\pi\xi)-\frac{\pi}{2}\biggr)\sum_{i=1}^3A_i^2
\nonumber\\[2mm]
&-&\frac{1}{\xi}\sum_{i\neq j}\frac{A_iA_j}{|\sin(2\gamma_{ij})|}\,
((\pi-|\gamma_{ij}|)\sinh(\xi|\gamma_{ij}|)-|\gamma_{ij}|\sinh(\xi(\pi-|\gamma_{ij}|)))\biggr\}\, .
\en
Using the asymptotic behavior for $R\to\infty$ of the radial wave function
\eq
f(R)\sim\sqrt{\frac{\pi}{2\kappa}}\,\exp(-\kappa R)\, ,
\en
and calculating, as in Ref.~\cite{MRR}, the asymptotic form of the second wave function
in Eq.~(\ref{eq:Eshift}) as $L\to\infty$, we arrive at the following expression for
the energy shift:
\eq
\Delta E&=&6\sqrt{3}\,N_{000}\,\sqrt{\frac{\pi}{2\kappa}}\,L^{-3/2}\sinh\biggl(\frac{\xi\pi}{2}\biggr)
\nonumber\\[2mm]
&\times&\biggl\{\sum_i\frac{A_i\exp(-\mu_{i(jk)}\kappa L)}{(\mu_{i(jk)})^{3/2}}\,
\int\frac{d^3{\bf x}_id^3{\bf y}_i}{|{\bf x}_i|}\,
(\psi^j({\bf x}_j,{\bf y}_j))^*V_j({\bf x}_j)
\nonumber\\[2mm]
&\times&\exp\biggl(\frac{\kappa\mu_{ki}}{\mu_{i(jk)}}\,{\bf x}_j{\bf e}
-\frac{\kappa\mu_{j(ki)}m_i}{\mu_{i(jk)}(m_i+m_k)}\,{\bf y}_j{\bf e}\biggr)
\nonumber\\[2mm]
&+&\biggl\{\sum_i\frac{A_i\exp(-\mu_{i(jk)}\kappa L)}{(\mu_{i(jk)})^{3/2}}\,
\int\frac{d^3{\bf x}_id^3{\bf y}_i}{|{\bf x}_i|}\,
(\psi^k({\bf x}_k,{\bf y}_k))^*V_k({\bf x}_k)
\nonumber\\[2mm]
&\times&\exp\biggl(\frac{\kappa\mu_{ij}}{\mu_{i(jk)}}\,{\bf x}_k{\bf e}
+\frac{\kappa\mu_{j(ki)}m_i}{\mu_{i(jk)}(m_i+m_j)}\,{\bf y}_k{\bf e}\biggr)\biggr\}\, .
\en
Using Eq.~(\ref{eq:Vpsi0}) and the normalization condition, we finally arrive at the following expression for the energy shift:
\eq\label{eq:l0}
\frac{\Delta E}{E_T}&=&-288\pi^2\sqrt{\frac{\pi}{2}}c_0
\sinh^2\biggl(\frac{\pi\xi}{2}\biggr)
(\kappa L)^{-3/2}
\nonumber\\[2mm]
&\times&\sum_{i\neq j}\exp(-\mu_{i(jk)}\kappa L)
\frac{A_iA_j}{(\mu_{i(jk)})^{3/2}}
\frac{I(|\gamma_{ij}|)}{|\sin(2\gamma_{ij})|}\, ,
\en
where
\eq
I(|\gamma_{ij}|)=\frac{\pi}{\xi\sinh(\pi\xi)}\,(\cosh(\xi(\pi-|\gamma_{ij}|))-\cosh(\xi|\gamma_{ij}|))\, .
\en
It can be checked that, in the equal mass limit, where $A_1=A_2=A_3=1/\sqrt{3}$,
the above formulae reduces to the result of Ref.~\cite{MRR} with the asymptotic
normalization  coefficient $A=1$. For illustrative purpose, one may rewrite Eq.~(\ref{eq:l0}) as
\eq\label{eq:Cz}
\frac{\Delta E}{E_T}=-(\kappa L)^{-3/2}\sum_{i=1}^3C_i\exp(-\mu_{i(jk)}\kappa L)\, ,
\en
where the coefficients $C_i$ depend on the masses in the system, but not on $L$
and the binding energy. In Fig.~\ref{fig:Cz} we plot the coefficients $C_1$ and
$C_2=C_3$ for a particular choice of the masses: $m_2=m_3$ and
$m_1/m_2=m_1/m_3=z$. As can be seen, at $z=1$, all $C_i$ are equal to $96.351\ldots/3
=32.117\ldots$ (cf. with Ref.~\cite{MRR}).

\subsection{The case $l=1$}

The wave function with $l_x=0$ and $l_y=l=1$ is given by
\eq
\psi^i_{1m}({\bf x}_i,{\bf y}_i)=\sum_{i=1}^3\phi^i_{1m}({\bf x}_i,{\bf y}_i)\, ,
\en
\eq
\phi^i_{1m}({\bf x}_i,{\bf y}_i)=N_{011}R^{-5/2}f(R)A_i\phi_1(\alpha_i)
\sqrt{\frac{4\pi}{3}}\,Y_{1m}(\Omega_{y_i})\, ,
\en
where
\eq
\phi_1(\alpha)&\!\!\!=\!\!\!&\frac{1}{2\sin(2\alpha)\cos\alpha}\,
\biggl(\sinh\biggl(\xi\biggl(\frac{\pi}{2}-\alpha\biggr)\biggr)\sin\alpha
-\xi\cosh\biggl(\xi\biggl(\frac{\pi}{2}-\alpha\biggr)\biggr)\cos\alpha\biggr)\, .
\en
It can be checked that the wave function obeys the equation
\eq\label{eq:Vpsi1}
V({\bf x}_j)\psi^j_{1m}({\bf x}_j,{\bf y}_j)
=-\delta^{(3)}({\bf x}_j)F_1({\bf y}_j)\, ,
\en
where
\eq\label{eq:F1}
F_1({\bf y}_j)=-\frac{\pi\xi A_j}{2M}\cosh\biggl(\frac{\xi\pi}{2}\biggr)\,
N_{011}\frac{K_{i\xi}(\kappa|{\bf y}_j|)}{|{\bf y}_j|}
\sqrt{\frac{4\pi}{3}}Y_{1m}(\Omega_{y_j})\, .
\en
Next, we consider the normalization condition. Here, we have to deal with the angular
integrations of two types. First, there are ``diagonal'' terms
\eq
\int d^3{\bf x}_id^3{\bf y}_i H(R,\alpha_i)Y^*_{1m}(\Omega_{y_i})
Y_{1m'}^{}(\Omega_{y_i})\, ,
\en
where $H(R,\alpha_i)$ denotes some function of the arguments $R$ and $\alpha_i$.
Using Eq.~(\ref{eq:hyperspherical}), it is immediately seen that the angular integrations
yield the factor $4\pi\delta_{mm'}$.
The ``non-diagonal'' terms have the following structure
\eq\label{eq:nondiag}
\int d^3{\bf x}_id^3{\bf y}_i \tilde H(R,\alpha_i,\alpha_j)Y^*_{1m}(\Omega_{y_i})
Y_{1m'}^{}(\Omega_{y_j})\, ,
\en
with some other function $\tilde H(R,\alpha_i,\alpha_j)$.
Using Eq.~(\ref{eq:ij}), it can be shown that
\eq\label{eq:Ytransform}
Y_{1m'}(\Omega_{y_j})=\frac{|{\bf x}_i|}{|{\bf y}_j|}\,(-\sin\gamma_{ij})
Y_{1m'}(\Omega_{x_i})+\frac{|{\bf y}_i|}{|{\bf y}_j|}\,(-\cos\gamma_{ij})
Y_{1m'}(\Omega_{y_i})\, .
\en
Performing the angular integrations, one should take into account the fact that,
owing to Eq.~(\ref{eq:costheta}), the variable $\alpha_j$ depends on the orientation
of both ${\bf x}_i$ and ${\bf y}_i$. Using this equation, the integral over $d\cos\theta$ can
be transformed into an integral over $\alpha_j$. The limits on the variation
of $\alpha_j$ are given by
\eq
||\gamma_{ij}|-\alpha_i|
\leq \alpha_j\leq
\frac{\pi}{2}-\biggl|\frac{\pi}{2}-\alpha_i-|\gamma_{ij}|\biggr|\, .
\en
Finally, the normalization condition takes the form
\eq
N_{011}^2=\kappa^2c_1\, ,
\en
where
\eq
c_1^{-1}=\frac{\pi\xi}{2\sinh(\pi\xi)}\,\sum_{i,j=1}^3A_iA_jI_{ij}\, .
\en
The diagonal terms can now be written as
\eq\label{eq:diag}
I_{ii}&\!\!\!=\!\!\!&\frac{\pi^2}{3}\int_0^{\pi/2}\frac{d\alpha}{\cos^2\alpha}\,\biggl(\sinh\biggl(\xi\biggl(\frac{\pi}{2}-\alpha\biggr)\biggr)\sin\alpha
-\xi\cosh\biggl(\xi\biggl(\frac{\pi}{2}-\alpha\biggr)\biggr)\cos\alpha\biggr)^2\, ,
\en
and the non-diagonal terms are given by
\eq\label{eq:nondiag1}
I_{ij}&=&-\frac{\pi^2}{3\,|\sin\gamma_{ij}|\cos^2\gamma_{ij}}
\int_0^{\pi/2}\frac{d\alpha\,\sin^2\alpha}{\sin^2(2\alpha)}
\nonumber\\[2mm]
&\times&\biggl(\sinh\biggl(\xi\biggl(\frac{\pi}{2}-\alpha\biggr)\biggr)\sin\alpha
-\xi\cosh\biggl(\xi\biggl(\frac{\pi}{2}-\alpha\biggr)\biggr)\cos\alpha\biggr)J_{ij}(\alpha)\, ,
\en
where
\eq
J_{ij}(\alpha)&\!\!\!\!=\!\!\!\!&\int_{\alpha_{min}}^{\alpha_{max}}
\frac{d\alpha'}{\cos^2\alpha'}\,
\biggl(\sinh\biggl(\xi\biggl(\frac{\pi}{2}-\alpha'\biggr)\biggr)\sin\alpha'
-\xi\cosh\biggl(\xi\biggl(\frac{\pi}{2}-\alpha'\biggr)\biggr)\cos\alpha'\biggr)
\nonumber\\[2mm]
&\!\!\!\!\times\!\!\!\!&
(\cos^2\gamma_{ij}+\cos^2\alpha-\sin^2\alpha')
\en
and
\eq
\alpha_{min}=||\gamma_{ij}|-\alpha|\, ,\quad\quad
\alpha_{max}=\frac{\pi}{2}-\biggl|\frac{\pi}{2}-\alpha-|\gamma_{ij}|\biggr|\, .
\en
Finally, the energy shift, averaged over all values of $m$, is given by\footnote{Note that, in higher partial waves, the energy shift depends on $m$ in the two-body bound states as well, see, e.g.,~Ref.~\cite{Hammer-Koenig}.}
\eq\label{eq:E-Pwave}
\frac{\Delta E}{E_T}&=&
-2\pi^2\sqrt{\frac{\pi}{2}}\xi^2\cosh^2\biggl(\frac{\xi\pi}{2}\biggr)c_1
(\kappa L)^{-3/2}
\nonumber\\[2mm]
&\times&\sum_{i\neq j}\exp(-\mu_{i(jk)}\kappa L)
\frac{A_iA_j}{(\mu_{i(jk)})^{3/2}}
\frac{1}{\sin\gamma_{ji}}\,
T(\cos\gamma_{ji})\, ,
\en
where
\eq
T(\alpha)=\frac{1}{\alpha}\,\int_0^\infty
K_{i\xi}\biggl(\frac{y}{\alpha}\biggr)\frac{d}{dy}\biggl(\frac{\sinh y}{y}\biggr)\, .
\en

\subsection{Arbitrary $l$}

The wave function in case of arbitrary $l$ is given by Eq.~(\ref{eq:philm})
with $l_x=0$ and $l_y=l$ (i.e., the resonant interaction is in the S-wave).
We can write this expression as
\eq\label{eq:philm1}
\phi^i_{lm}({\bf x}_i,{\bf y}_i)=N_{0ll}R^{-5/2}f(R)A_i\phi_l(\alpha_i)
\sqrt{\frac{4\pi}{2l+1}}\,Y_{lm}(\Omega_{y_i})\, ,
\en
where the Jacobi functions, entering this expression, can be determined from
certain recurrence relations. These relations can be obtained
from the definition of the Jacobi functions
\eq
P_\nu^{a,b}(x)=\frac{\Gamma(\nu+a+1)}{\Gamma(\nu+1)\Gamma(a+1)}\,
F\biggl(-\nu,\nu+a+b+1,a+1,\frac{1}{2}\,(1-x)\biggr),
\en
as well as the
recurrence relations for the hypergeometric functions $F$, see,
e.g., Ref.~\cite{Gradsteyn}. The recurrence relations for the Jacobi functions take
the form
\eq
\biggl(\nu+\frac{a+b}{2}+1\biggr)(1-x)P_\nu^{a+1,b}(x)&=&(\nu+a+1)P_\nu^{a,b}(x)
-(\nu+1)P_{\nu+1}^{a,b}(x)\, ,
\nonumber\\[2mm]
\biggl(\nu+\frac{a+b}{2}+1\biggr)(1+x)P_\nu^{a,b+1}(x)&=&(\nu+b+1)P_\nu^{a,b}(x)
+(\nu+1)P_{\nu+1}^{a,b}(x)\, ,
\en
starting from
\eq
P_\nu^{1/2,1/2}(\cos2\alpha)=\frac{\Gamma(\nu+3/2)}{\Gamma(\nu+1)\Gamma(3/2)}\,
\frac{\sin(2(\nu+1)\alpha)}{(\nu+1)\sin 2\alpha}\, .
\en
Substituting the expression for the wave function into the normalization condition, the
diagonal integral (analog of Eq.~(\ref{eq:diag})) reads
\eq
I_{ii}=\frac{16\pi^2}{2l+1}\,\int_0^{\pi/2}d\alpha\sin^2\alpha\cos^2\alpha(\phi_l(\alpha))^2\, ,
\en
whereas the non-diagonal integral (analog of Eq.~(\ref{eq:nondiag1})) is given by
\eq
I_{ij}=\frac{4\pi}{2l+1}\,\int d\Omega_{x_i}d\Omega_{y_i} d\alpha_i
\sin^2\alpha_i\cos^2\alpha_i\phi_l(\alpha_i)\phi_l(\alpha_j)
Y^*_{lm}(\Omega_{y_i})Y_{lm}(\Omega_{y_j})\, .
\en
In general, the transformation between the wave functions, depending on different
sets of Jacobi coordinates, is given by the Raynal-Revai coefficients~\cite{Raynal:1970ah}.
An explicit expression for these coefficients is known in the literature (see, e.g.,
Ref.~\cite{Ershov:2017dlp} and earlier references therein).
However here we do not make use of these rather voluminous formulae.
Rather, in order to calculate the angular integral, in analogy with Eq.~(\ref{eq:nondiag}),
we express the quantity $Y_{lm}(\Omega_{y_j})$
as a sum of products $Y_{l'm'}(\Omega_{y_i})Y_{l''m''}(\Omega_{x_i})$
with all possible $l'+l''\leq l$ and $m'+m''=m$. In order to do this, is it useful to define
the solid harmonics:
\eq
{\cal Y}_{lm}({\bf y}_j)=|{\bf y}_j|^lY_{lm}(\Omega_{y_j})\, .
\en
The quantity ${\cal Y}_{lm}({\bf y}_j)$ is a polynomial of power $l$ in the components
of the 3-vector ${\bf y}_j$. Writing  ${\bf y}_j=a{\bf y}_i+b{\bf x}_i$,
one immediately sees that each term in the expression of ${\cal Y}_{lm}({\bf y}_j)$
decomposes into  monomials of the components of the vectors ${\bf y}_i$ and ${\bf x}_i$
of power $l_1$ and $l_2$, respectively, with $l_1+l_2=l$. These monomials, in their
turn, can be expressed through  ${\cal Y}_{l'm'}({\bf y}_i)$ and
${\cal Y}_{l''m''}({\bf x}_i)$, respectively, with $l'\leq l_1$ and $l''\leq l_2$, leading to the
above-mentioned expansion.

Further, one has to calculate integrals of the type
\eq
I_\Omega=\int d\Omega_{x_i}d\Omega_{y_i}\phi_l(\alpha_j)
Y^*_{lm}(\Omega_{y_i})Y_{l'm'}(\Omega_{y_i})Y_{l''m''}(\Omega_{x_i})\, .
\en
Let us recall here that $\alpha_j$
depends on the scalar product ${\bf x}_i{\bf y}_i$, so the two angular integrations do
not immediately decouple. In order to achieve this decoupling, consider first the integration
over $d\Omega_{y_i}$, with the direction of the unit vector $\hat x_i$ fixed. Note that it is always
possible to find a rotation $R_{\bf x}$ so that
\eq
R_{\bf x}\hat x_i={\bf e}\, ,\quad\quad {\bf e}=(0,0,1)\, .
\en
Perform now the variable transformation $y_i=R_{\bf x}^{-1}y'_i$, with
$d\Omega_{y_i}=d\Omega'_{y_i}$. After this transformation, we have
${\bf x}_i{\bf y}_i={\bf e}{\bf y}_i'$. Further,
\eq
Y_{lm}(\Omega_{y_i})&=&\sum_{n=-l}^lD^{(l)}_{mn}(R_{\bf x}^{-1})Y_{ln}(\Omega'_{y_i})\, ,
\nonumber\\[2mm]
Y_{l'm'}(\Omega_{y_i})&=&\sum_{n'=-l'}^{l'}D^{(l')}_{m'n'}(R_{\bf x}^{-1})Y_{l'n'}(\Omega'_{y_i})\, ,
\nonumber\\[2mm]
Y_{l''m''}(\Omega_{x_i})&=&\sum_{n''=-l''}^{l''}D^{(l'')}_{m''n''}(R_{\bf x}^{-1})Y_{l''n''}(\Omega_e)\, ,
\en
where the $D^{(l)}$ denote Wigner $D$-matrices in the irreducible representation of
the rotation group, characterized by the angular momentum $l$. It is now seen that
the integration over two solid angles decouple:
\eq
I_\Omega&=&\sum_{nn'n''}\int d\Omega_{x_i}(D^{(l)}_{mn}(R_{\bf x}^{-1}))^*
D^{(l')}_{m'n'}(R_{\bf x}^{-1})D^{(l'')}_{m''n''}(R_{\bf x}^{-1})
\nonumber\\[2mm]
&\times&\int d\Omega'_{y_i}\phi_l(\alpha_j)
Y^*_{lm}(\Omega_{y'_i})Y_{l'm'}(\Omega_{y'_i})Y_{l''m''}(\Omega_{n})\, .
\en
Here, the quantity $\alpha_i$ is determined by Eq.~(\ref{eq:alphaij}) with
$\theta_i$ denoting the angle between the unit vectors $\hat y'_i$ and ${\bf e}$, so that
$\cos\theta_i=\cos\theta$, $d\Omega'_{y_i}=d\cos\theta d\varphi$ and $Y_{lm}(\Omega'_{y_i})=Y_{lm}(\theta,\varphi)$. The integral over $d\Omega_{x_i}$ can be finally performed,
yielding a group-theoretical factor, and one is left only with the integral over the
solid angle $d\Omega'_{y_i}$.
It does not make much sense to present the (quite voluminous)
general result here. If needed, it can be straightforwardly derived in each particular
case along the lines described above.

Next, one needs an analog of Eqs.~(\ref{eq:Vpsi0}), (\ref{eq:F0})
and Eqs.~(\ref{eq:Vpsi1}), (\ref{eq:F1}) in case of arbitrary $l$.
To this end, using the explicit form of $\phi_l(\alpha)$,
it suffices to represent the wave function $\phi^i_{lm}({\bf x}_i,{\bf y}_i)$ in
Eq.~(\ref{eq:philm1}) as
\eq
\phi^j_{lm}({\bf x}_j,{\bf y}_j)=\frac{1}{4\pi |{\bf x}_j|}\,F_l({\bf y}_j)
+\tilde \phi^j_{lm}({\bf x}_j,{\bf y}_j)\, ,
\en
where the second term on the right-hand side is regular as $|{\bf x}_i|\to 0$.
Then, the analog of Eqs.~(\ref{eq:Vpsi0}), (\ref{eq:Vpsi0}) reads
\eq
V_j({\bf x}_j)\psi^j({\bf x}_j,{\bf y}_j)=-\delta^{(3)}({\bf x}_j)F_l({\bf y}_j)\, .
\en
With these building blocks, the leading contribution to
the energy shift expression can be straightforwardly calculated
\eq\label{eq:DeltaEm}
\Delta E_m&=&-\sum_{i=1}^3\sum_{\bf e}\int d^3{\bf x}_id^3{\bf y}_i\delta^3({\bf x}_i)
(F_l({\bf y}_i))^*
\nonumber\\[2mm]
&\times&\biggl(\phi^j_{lm}({\bf x}_j,{\bf y}_j+{\bf e}L\mu_{j(ki)})
+\phi^k_{lm}({\bf x}_k,{\bf y}_k-{\bf e}L\mu_{k(ij)})\biggr)\, .
\en
Here, we take into account the fact that the finite-volume energy shift can explicitly
depend on the projection of the angular momentum $m$.

In order to proceed further, we note that, for arbitrary $l$, the function
$\phi_l(\alpha)$ is singular at $\alpha=0$:
\eq
\phi_l(\alpha)=\frac{G_l}{\alpha}+\tilde\phi_l(\alpha)\, ,
\en
where the second term is regular at the origin. The leading contribution in the limit
$L\to\infty$ comes from the singular term. Further, in this limit, we have
\eq
\lim_{L\to\infty}Y_{lm}(\Omega_{y_j'})=Y_{lm}(\Omega_e)\, ,\quad\quad
\lim_{L\to\infty}Y_{lm}(\Omega_{y_k''})=(-1)^lY_{lm}(\Omega_e)\, ,
\en
where ${\bf y}_j'={\bf y}_j+{\bf e}L\mu_{j(ki)}$ and
${\bf y}_k''={\bf y}_k-{\bf e}L\mu_{k(ij)}$.

In the following, we present the averaged shift, defined as
\eq
\Delta E=\frac{1}{2l+1}\,\sum_{m=-l}^l\Delta E_m\, .
\en
Defining $F_l({\bf y}_i)=\bar F_l(|{\bf y}|_i)Y_{lm}(\Omega_{y_i})$,
Eq.~(\ref{eq:DeltaEm}) can be finally transformed into
\eq
\Delta E&=&-3\biggl(\frac{4\pi}{2l+1}\biggr)^{1/2}
\biggl(\frac{\pi}{2\kappa}\biggr)^{1/2}\sum_{i=1}^{3}
N_{0ll}G_l\int_{-1}^{1}dzP_l(z)\int_0^\infty ydy(\bar F(y))^*
\nonumber\\[2mm]
&\times&\biggl(\frac{A_j(L\mu_{j(ki)})^{-3/2}}{|\sin\gamma_{ij}|}\,
\exp(-\kappa L\mu_{j(ki)})\exp(\kappa\cos\gamma_{ij}yz)
\nonumber\\[2mm]
&+&(-1)^l\frac{A_k(L\mu_{k(ij)})^{-3/2}}{|\sin\gamma_{ik}|}\,
\exp(-\kappa L\mu_{k(ij)})\exp(\kappa\cos\gamma_{ik}yz)\biggr)\, .
\en
From the above expression, it is clear that the result for general $l$
looks similar to Eqs.~(\ref{eq:Cz}), (\ref{eq:E-Pwave}). Namely, it contains
the exponentially vanishing factors together with an overall factor
$(\kappa L)^{-3/2}$. Only the numerical coefficients depend on the
angular momentum $l$.

\section{Conclusions}

\begin{itemize}

\item[i)]
In this article, we have extended the approach of Ref.~\cite{MRR} and derived
explicit expressions for the energy shift of the three-particle bound state
in the unitary limit  with non-equal mass constituents and with the
total angular momentum different from zero. All cases of physically relevant
angular momenta (i.e., for which the the shallow bound states exist in the unitary limit) were covered.

\item[ii)]
We show that the behavior of the leading terms in the finite-volume energy shift
is universal for all $l$: namely, it contains three exponentially vanishing
 terms, whose arguments are determined by the pertinent reduced masses, i.e.,
by pure kinematics. In addition, there is a common multiplicative
factor $(\kappa L)^{-3/2}$
for all $l$. Only the numerical coefficients, which stand in front of these
universal factors, depend on $l$, and can be calculated for each $l$ explicitly,
using the method described in the paper.

\item[iii)]
On several occasions already, the simple model, considered in
Ref.~\cite{MRR}, has
served as a nice testing ground for the different types of the
three-particle quantization condition, which are available in the literature
(see, e.g., \cite{Hansen-corr,Pang1}). Moreover, a comparison of the results
has shed more light on the role of a three-particle force in the description of
the volume-dependence of the shallow bound states~\cite{Pang1}.
A universal formula for arbitrary $l$
and unequal masses, which was derived in this paper, without any doubt,
represents a further
challenge for the above-mentioned approaches, as well as an opportunity
to gain a deeper insight in the three-particle dynamics in a finite volume.

\end{itemize}

\subsection*{Acknowledgments}

The authors thank H.-W. Hammer for useful discussions.
 We acknowledge the support from the CRC 110
``Symmetries and the Emergence of Structure in QCD''
(DFG grant no. TRR~110 and NSFC grant No. 11621131001).
This research is supported in part by Volkswagenstiftung under contract no. 93562,
by the Chinese Academy of Sciences (CAS) President's International Fellowship
Initiative (PIFI) (Grant No.\! 2018DM0034)
and by Shota Rustaveli National Science Foundation (SRNSF), grant no. DI-2016-26.
It is also supported in part by the National Science Foundation of China (NSFC) under
the  project  No.11335001 and by Ministry of Science and Technology of
China (MSTC) under 973 project "Systematic studies on light hadron spectroscopy", No. 2015CB856702.

\end{document}